# Variable Transmission Voltage for Loss Minimization in Long Offshore Wind Farm AC Export Cables

Bjørn Gustavsen, *Fellow, IEEE*, and Olve Mo

*Abstract*—Connection of offshore wind farms to shore requires the use of submarine cables. In the case of long HVAC connections, the capacitive charging currents limit the transfer capability and lead to high losses. This paper shows that the losses can be substantially reduced by continuously adjusting the cable operating voltage according to the instantaneous wind farm power production. Calculations for a 320 MW windfarm connected to shore via a 200 km cable at 220 kV nominal voltage shows that an annual loss reduction of 9% is achievable by simply using a ±15% tap changer voltage regulation on the two transformers. Allowing a larger voltage regulation range leads to further loss reduction (13% for 0.4-1.0 p.u. voltage range). If the windfarm has a low utilization factor, the loss reduction potential is demonstrated to be as high as 21%. The methodology can be applied without introducing new technology that needs to be developed or qualified.

*Index Terms*— Wind energy, wind farm, offshore wind, submarine cable, power engineering computing, export cable, cable connection, operation, voltage control, losses, optimization.

## I. INTRODUCTION

Most offshore wind farms are connected to the onshore grid via HVAC cables. As it is often desirable to locate the wind farm at long distances from shore, e.g. due to more favorable wind conditions, the increased losses in the HVAC cables that result from charging currents can make the development of the wind farm economically or even technically infeasible. This limitation for long HVAC cables has motivated the use of HVDC connections to shore. The HVDC solution gives lower losses but represents a step in investment and operating cost [1]-[5].

The increased cost and complexity of HVDC solutions has motivated a search for methods to extend the feasibility of the HVAC alternative. One possibility is to introduce reactive shunt compensation at one or more positions along the cable but such solution requires either sub-sea compensation equipment or additional offshore platforms. Another alternative is to use a lower frequency than the standard 50/60 Hz frequency [6], thereby reducing both charging currents and skin effect in the conductors. One important disadvantage of the low-frequency AC alternative is the need of specialized components that has not yet been qualified for this use. Other disadvantages are the increased weight and volume of the magnetic components as well as the need for an onshore converter station.

In this work, we propose an alternative solution which is entirely based on existing 50/60 Hz AC technology and which does not require the use of additional reactive compensation along the cable route. This solution is motivated by the observation that the cable losses associated with the cable charging currents decrease with decreasing voltage. That way, the total losses in the cable can be reduced when the wind farm production is low since the losses associated with charging currents may dominate over those of the transmitted power. By varying the operating voltage of the export cables by transformer on-line tap changers, we show that it is possible to both reduce the cable losses and to extend the technical range limits of the AC cable. Two strategies are investigated. 1) Operating at a fixed, optimized voltage, or 2) operating at variable voltage that is continuously optimized for the instantaneous wind farm production. The two operating strategies are compared using wind farm power production and cable length as parameters. Finally, the potential reduction in annual losses are determined by considering the distribution of the wind farm production over one year of operation. The results are shown for two alternative distribution profiles.

## II. SYSTEM OVERVIEW

This study considers a system consisting of an aggregated wind farm that is connected to land grid via an AC cable as shown in Fig. 1. The cable operating voltage is controlled by the interfacing transformers that are assumed to have on-line tap-changers. The tap-changer is adjusted based on the wind farm active power production. Reactive compensation is provided by the indicated shunt reactors and/or the onshore grid and wind farm.

The study is based on a 220 kV cable whose electrical parameters and current rating are listed in Table I [2].

B. Gustavsen and O. Mo are with SINTEF Energy Research, N-7465 Trondheim, Norway (e-mail: bjorn.gustavsen@sintef.no, olve.mo@sintef.no).
This work was supported by the Norwegian Research Centre for Offshore Wind Technology (Nowitech). https://www.sintef.no/projectweb/nowitech/

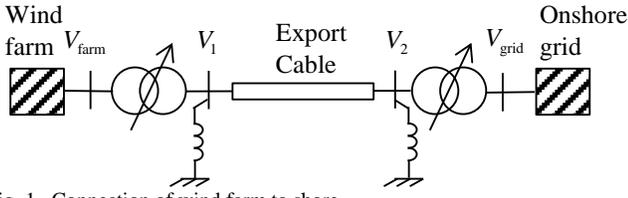

Fig. 1. Connection of wind farm to shore.

TABLE I.
CABLE PARAMETERS (50 HZ) AND CHARACTERISTICS [2].

| Nominal voltage | 220 kV |
|---|---|
| Cable section [mm$^2$] | 1000 |
| $R$ [Ω/km] | 0.048 |
| $L$ [mH/km] | 0.37 |
| $C$ [μF/km] | 0.18 |
| $G$ [S/km] | 0 |
| Nominal current [A] | 1055 |

## III. SYSTEM LEVEL MODELING

The system in Fig. 1 is represented by the electrical circuit depicted in Fig. 2 that is assumed to operate at 50 Hz. An aggregated representation by voltage sources is used for the systems on the grid-side and offshore side of the cable.

- The onshore grid voltage is assumed fixed and equal to its nominal value (e.g. 380 kV).
- The grid connection transformer is assumed ideal with voltage ratio $k$. The cable onshore voltage is therefore

$$V_2 = k \cdot V_{\text{grid}} \qquad (1)$$

- The onshore grid and the onshore transformer with tap-changer are modelled by the ideal voltage source $V_2$. The effect of the tap-changer is represented by varying the amplitude of the voltage source $V_2$.
- The wind farm connection transformer is also assumed to be ideal. The voltage on the cable side is allowed to exceed the cable voltage on the cable onshore side by 10%. The permissible operating area is specified as

$$V_1 = V_2 \cdot \alpha \cdot e^{j\beta} \qquad (2)$$

where $\alpha \in [1, 1.1]$

- It is assumed that the reactive power consumption can be controlled such that the cable wind side voltage ($V_1$) in (2) is within the permissible range. The voltage source $V_1$ represents the aggregated effect of tap-changer, transformer, wind turbine converters and reactive power compensation equipment.
- The cable is represented by its exact PI-equivalent, accurately taking into account the distributed parameter effects and thereby the variation of voltage and current along the cable. The details are shown in Section IV.

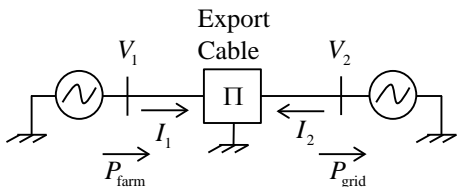

Fig. 2. Electrical equivalent with cable represented by exact pi-equivalent.

## IV. CABLE MODELING

### A. Cable Terminal Admittance Matrix

The cable behavior is defined by its length $l$, its per-unit-length (PUL) series impedance $Z$ and shunt admittance $Y$,

$$Z(\omega) = R + j\omega L \qquad (3)$$
$$Y(\omega) = G + j\omega C \qquad (4)$$

From the PUL parameters, the cable admittance matrix is obtained as [7]

$$\mathbf{Y}_\Pi(\omega) = \begin{bmatrix} \dfrac{\coth(\gamma(\omega)l)}{Z_C(\omega)} & \dfrac{-1}{Z_C(\omega)\sinh(\gamma(\omega)l)} \\ \dfrac{-1}{Z_C(\omega)\sinh(\gamma(\omega)l)} & \dfrac{\coth(\gamma(\omega)l)}{Z_C(\omega)} \end{bmatrix} \qquad (5)$$

where

$$Z_C(\omega) = \sqrt{\dfrac{Z(\omega)}{Y(\omega)}} \qquad (6)$$

$$\gamma(\omega) = \sqrt{Z(\omega)Y(\omega)} \qquad (7)$$

Using the admittance matrix (5) together with the (known) terminal voltages, the cable terminal currents are calculated as

$$\begin{bmatrix} I_1 & I_2 \end{bmatrix}^T = \mathbf{Y}_\Pi \begin{bmatrix} V_1 & V_2 \end{bmatrix}^T \qquad (8)$$

### B. Cable Loss Calculation

From the solution of currents at the cable ends, the cable active and reactive power transmitted from the wind farm (farm) and absorbed at the land side (grid) are calculated as

$$P_{\text{farm}} = \sqrt{3}\,\text{Re}\{V_1 I_1^*\}, \quad Q_{\text{farm}} = \sqrt{3}\,\text{Im}\{V_1 I_1^*\} \qquad (9a)$$

$$P_{\text{grid}} = -\sqrt{3}\,\text{Re}\{V_2 I_2^*\}, \quad Q_{\text{grid}} = -\sqrt{3}\,\text{Im}\{V_2 I_2^*\} \qquad (9b)$$

and the cable losses are obtained as

$$P_{\text{loss}} = \sqrt{3}\,\text{Re}\{V_1 I_1^* + V_2 I_2^*\} \qquad (10)$$

### C. Cable Internal Voltages and Currents

In order to assess the voltage and current at ($N$–1) internal nodes, the cable is subdivided into $N$ segments of equal length $l_{\text{seg}} = l/N$ as shown in Fig. 3. The admittance matrix is calculated by (5) with length $l_{\text{seg}}$, and the global admittance matrix is assembled using nodal analysis. With the voltages at the two cable ends taken as known quantities, the internal voltages and currents are calculated using nodal analysis. This method is used for monitoring the voltage and current along the cable.

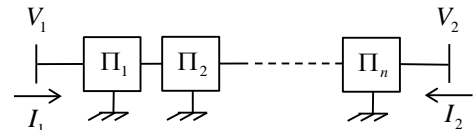

Fig. 3. Segmentation of cable into $n$ sections for assessment of internal voltages and currents.

## V. CABLE EFFICIENCY

### A. Definition

The objective is to operate the system in such way that the cable efficiency is maximized, defined as the ratio between transmitted power to the grid ($P_{\text{grid}}$) and the produced power at the wind farm ($P_{\text{farm}}$) as shown in (11).

$$\eta = \frac{P_{\text{grid}}}{P_{\text{farm}}} \quad (11)$$

It is remarked that maximal cable efficiency implies minimal cable losses since we have $P_{\text{loss}} = P_{\text{grid}} - P_{\text{farm}}$.

To analyze the efficiency, we start by rewriting (8) as

$$\begin{bmatrix} I_1 \\ I_2 \end{bmatrix} = \begin{bmatrix} A & B \\ B & A \end{bmatrix} \cdot \begin{bmatrix} \xi V_2 \\ V_2 \end{bmatrix} \quad (12)$$

where $\xi$ is the voltage scaling required for achieving a given active and reactive power flow,

$$\xi = \alpha \cdot e^{j\beta} \quad (13)$$

In (12), $A$, $B$ and $\xi$ are complex quantities while $V_2$ can be assumed real-valued. Combining (12) with the expressions for power in (9) gives

$$P_{farm} = \sqrt{3} \, \text{Re}\{\xi V_2 (A\xi V_2 + BV_2)^*\} \quad (14a)$$
$$= \sqrt{3} \, \text{Re}\{\xi (A\xi + B)^*\} V_2^2$$

$$P_{\text{grid}} = -\sqrt{3} \, \text{Re}\{V_2 (B\xi V_2 + AV_2)^*\} \quad (14b)$$
$$= -\sqrt{3} \, \text{Re}\{\xi (B\xi + A)^*\} V_2^2$$

and for the cable efficiency

$$\eta = \frac{P_{\text{grid}}}{P_{\text{farm}}} = -\frac{\text{Re}\{\xi (B\xi + A)^*\}}{\text{Re}\{\xi (A\xi + B)^*\}} \quad (15)$$

It is noted from (15) that the cable efficiency is independent of the operating voltage $V_2$ and that, consequently, the maximum achievable efficiency is also independent of the voltage $V_2$. Note however that the operating voltage $V_2$ affects the maximum power that can be transmitted and more importantly the power that is transmitted at maximum efficiency.

Fig. 4 shows the cable efficiency $\eta$ for the 220 kV cable with parameters as given in Table I. The efficiency is shown as function of the scale angle $\beta$ in (13) with unity scaling ($\alpha=1$), for alternative cable lengths. It is seen that for each cable length, there exists a unique scale angle that maximizes the efficiency. The maximum achievable efficiency decreases with the cable length. The operating voltage $V_2$ have no effect on the maximum achievable efficiency.

It is remarked that the longest cables can only be operated with a low voltage $V_2$ since the charging currents will otherwise cause the rated current to be exceeded at the cable ends. Therefore, the power transfer capability is very much reduced, making such lengths economical unfeasible. In this work, we will instead focus on the cable efficiency in the case of moderate cable lengths. The maximum transfer capability is studied separately in Section VIII-H.

### B. Optimal Efficiency

In the case of long cables, the cable efficiency can be further improved by also controlling the scaling factor $\alpha$, in addition to the angle $\beta$. Fig. 5 shows the efficiency $\eta$ as function of $\beta$ with the scaling $\alpha$ as parameter, for a cable length of 200 km. For the given cable and length, the efficiency can never exceed $\eta_{\text{opt}}=0.94$ which then represents a theoretical upper limit for this cable parameter set. From the results in Fig. 4 and 5, we can conclude that for a given cable type and length, there exists a scaling $\xi = \alpha \cdot e^{j\beta}$ that optimizes the cable operation in terms of cable efficiency. Both $\alpha$ and $\beta$ should therefore be used for controlling the cable operation, which is the principle used in this work. It is remarked that the optimum value for $\alpha$ approaches 1.0 as the cable length is reduced while $\beta$ approaches 0.

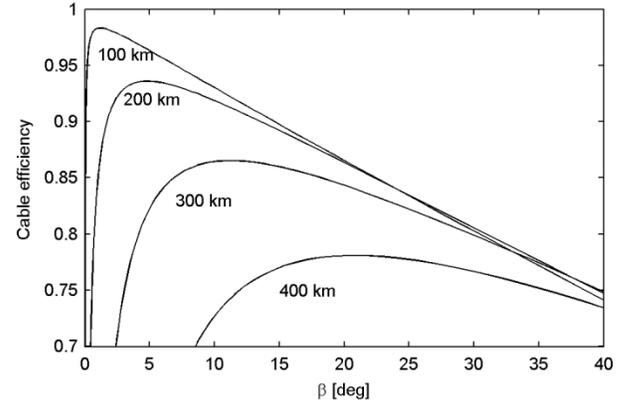

Fig. 4. Cable efficiency as function of wind farm voltage scaling, $\xi = 1 \cdot e^{j\beta}$ for different cable lengths.

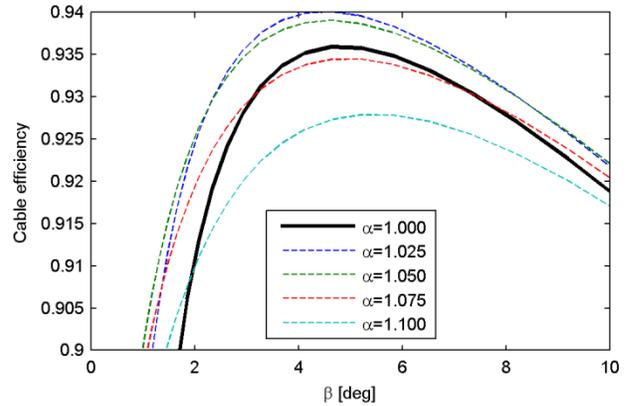

Fig. 5. Cable efficiency of 200 km cable as function of wind farm voltage scaling, $\xi = \alpha \cdot e^{j\beta}$

### C. Optimal Operating Voltage

In principle, one could for a given cable type and length simply identify the scaling $\xi_{\text{opt}}$ which corresponds to the best point in Fig. 5, and then establish a curve which relates the (optimum) operating voltage $V_2$ to the wind farm instantaneous power production,

$$V_{2,\text{opt}} = f(P_{\text{farm}}, \xi_{\text{opt}}). \qquad (16)$$

This curve is shown in Fig. 6 for the best point defined by $\alpha=1.025$, $\beta=4.25°$. It is observed that as the wind farm production increases, the maximum permissible operating voltage ($V_2=1.0$ p.u.) is exceeded at about 200 MW, and the current limit (1055 A) is exceeded at about 250 MW as indicated by the asterisk.

In order to operate the cable at such high power transfers, it therefore becomes necessary to modify the choice of operating voltage $V_2$ and voltage scaling $\xi$. In the next sections we will achieve this by searching for the combination of $V_2$, $\alpha$, and $\beta$ which satisfies the required production without exceeding the permissible limits on cable voltage and current, while maximizing the cable efficiency.

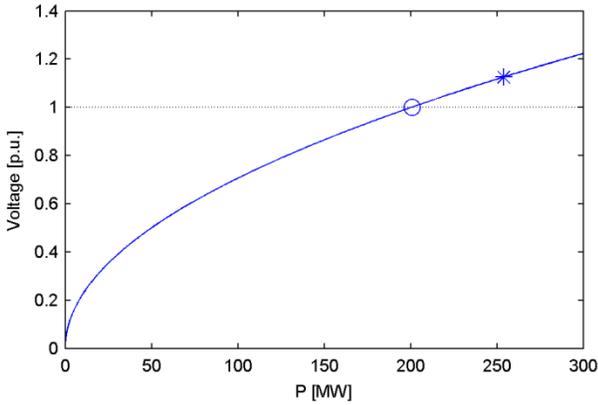

Fig. 6. Optimal cable operating voltage $V_2$ as function of wind farm instantaneous production $P_{wf}$, for maximum efficiency for the 200 km cable. The asterisk and circle denote the operating voltage $V_2$ at which the cable rated current and rated voltage are exceeded, respectively.

## VI. OPERATING STRATEGIES

### A. Fixed Transmission Voltage

One possible operating principle is to use a fixed voltage $V_2$ which may be lower than the nominal voltage.

Fig. 7 shows the cable efficiency as function of the wind farm production with alternative operating voltages $V_2$ in p.u. of the nominal voltage, for a cable length of 200 km. The results are shown up the point where the required power transfer becomes technically infeasible. Clearly, the fixed operating voltage should be chosen based on the expected production level, with lower voltage for low production levels.

### B. Variable Transmission Voltage

A better strategy is to operate the cable with a variable transmission voltage $V_2$ that is chosen based on the instantaneous wind production. In this case, it is necessary to determine the voltage that gives the lowest cable losses for each production level while not exceeding the voltage and current limits.

Fig. 8 shows the optimum cable voltage as function of the wind farm production, assuming that the operating voltage is permitted to vary in the range 0.4 p.u.-1.0 p.u. The result is shown for cable lengths 100 km, 200 km and 300 km. It is seen that the optimum voltage decreases as wind farm production is reduced, consistently with the result in Fig. 6.

The corresponding cable efficiency is shown in Fig. 9 with solid traces. For comparison, the result with 1.0 p.u. (fixed) operating voltage is shown with dashed traces. It is observed that use of a variable transmission voltage can greatly increase the cable efficiency in periods where wind farm production is low. In the case of the 300 km length, the dashed trace is missing since operation at 1.0 p.u. is not feasible.

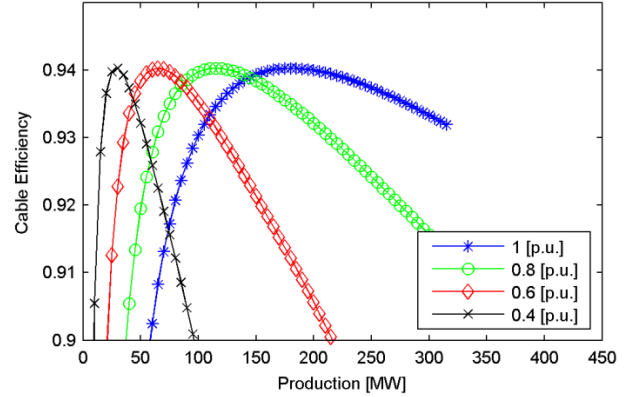

Fig. 7. Cable efficiency for 200 km cable as function of wind farm instantaneous active power production. Parameter: Cable operating voltage $V_2$.

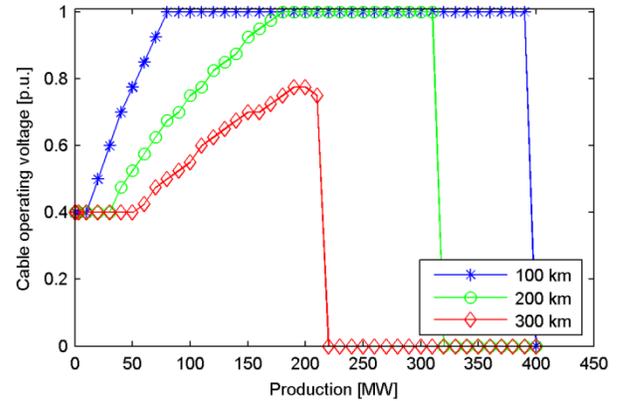

Fig. 8. Optimal cable operating voltage as function of wind farm instantaneous active power production. Parameter: cable length.

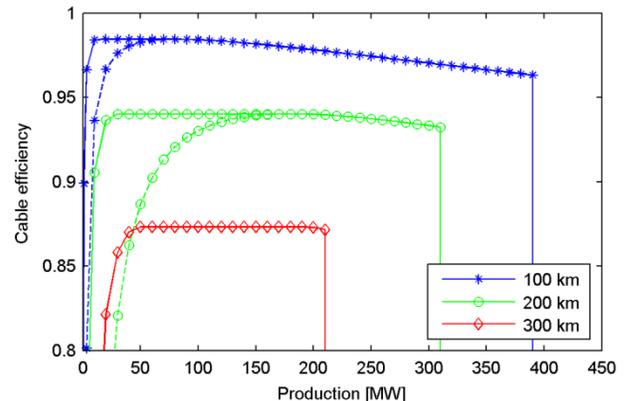

Fig. 9. Cable efficiency as function of wind farm instantaneous active power production. Solid lines: Operation at optimal (variable) voltage. Dashed lines: operation at 1.0 p.u. fixed voltage. Parameter: cable length.

## VII. LOSS MINIMIZATION WITH REPRESENTATIVE DISTRIBUTION OF ANNUAL WIND FARM PRODUCTION

### A. Wind Farm Annual Production

The advantages by use of a reduced, fixed operating voltage, or a variable operating voltage, are dependent on the wind farm production profile. In order to quantify the advantage we make use of the annual efficiency defined as

$$\eta_{\text{annual}} = \frac{\sum_{i=1}^{N} \Delta t_i P_{\text{grid},i}}{\sum_{i=1}^{N} \Delta t_i (P_{\text{farm},i} + P_{\text{curtail},i})} \quad (17)$$

The term $P_{\text{curtail},i}$ in (17) represents curtailment due to lack of cable capacity. Thus, any wind energy that is not produced due to lack of transfer capacity is treated as losses in the following calculations.

### B. Example: Wind Farm With High Utilization Factor

As an example, we consider the ten-year distribution of the power production of a windfarm. The example used is a synthesized power production for the NOWITECH reference wind farm [8]. This farm is considered representative for a wind farm at Doggerbank in the North Sea.

Fig. 10 shows the relative duration of the wind farm production with a resolution of $N=100$ points. The capacity utilization factor for this data set is 0.46, defined as the average annual energy production divided by the theoretical maximum annual production (rated production year around, no curtailment).

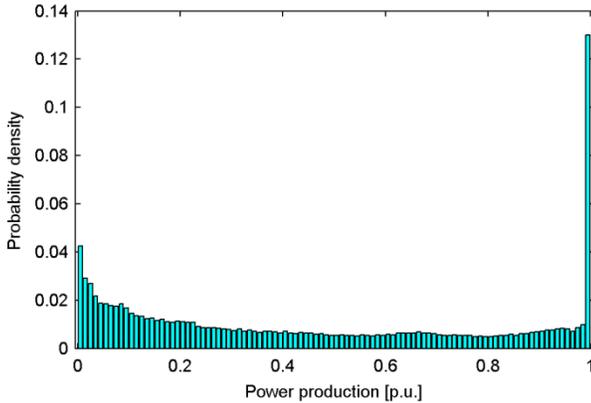

Fig. 10. Distribution of wind farm power production in [p.u.] of maximum installed windfarm production. Data for wind farm with high utilization factor.

Fig. 11 shows the annual cable efficiency calculated by (17) as function of the installed power at the wind farm, assuming that the distribution in Fig. 10 is independent of the installed production. The annual efficiency is shown for alternative operating conditions for the cable: Operating with a fixed voltage, or operating with a variable voltage in the range 0.4-1.0 p.u. It is observed that a fixed operating voltage should be chosen based on the installed power. The maximum achievable annual efficiency is anyhow limited to about 0.925. It is further observed that by allowing the voltage to vary in the range 0.4-1.0 p.u., the annual efficiency can be increased to nearly 0.94 for a wide range of installed powers, being close to the theoretical upper limit of 0.94 in Fig. 4. With $P_{\text{farm}}=320$ MW, the increase of annual efficiency is somewhat lower, from about 0.925 to 0.935 compared to operating at 1.0 p.u. fixed voltage. Still, this represents 13% reduction in losses.

### C. Loss Reduction Utilizing Tap Regulation

Transformers with online tap changers (OLTC) can be used to adapt the operating voltage to the instantaneous wind farm production. That way, the cable efficiency can be improved compared to the operation with fixed voltage shown in Fig. 11.

Fig. 12 shows the cable efficiency curves corresponding to those in Fig. 11 when assuming that the transformers have OLTC capability of ±15%. The result is shown with dashed traces when the nominal tap setting is 0.4, 0.6, or 0.87 p.u. The corresponding result with fixed voltage is shown with solid traces. Comparison between solid and dashed traces in Fig. 12 shows that utilizing the transformer voltage regulating capability of ±15% can improve the cable efficiency with almost 1%.

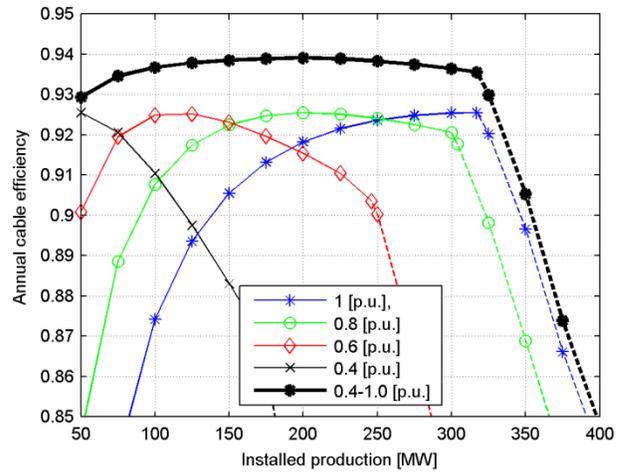

Fig. 11. Annual cable efficiency as function of wind farm maximum instantaneous production. The dashed portion of traces represents wind farms that can produce more than cable maximum capacity such that production curtailment will be required (causing the steep drop in annual efficiency)

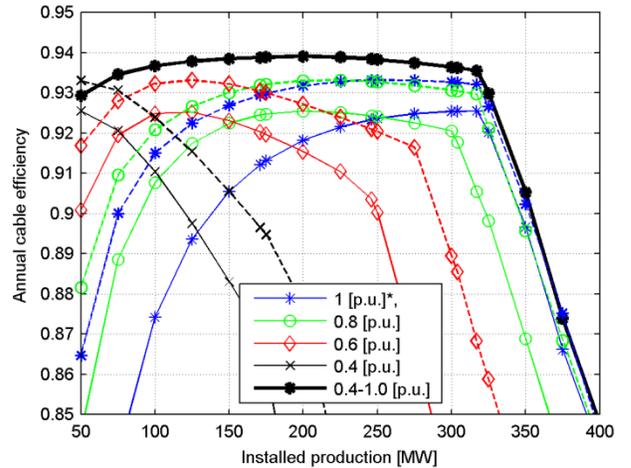

Fig. 12. Annual cable efficiency as function of wind farm maximum instantaneous production with cable operating voltage as parameter. The result with ±15% regulating capability is shown with dashed traces. The two traces with asterisk, (blue) are a special case for wich the solid trace is operation at 1.0 p.u. and the dashed trace is operation at 0.87 p.u ±15% such that the maximum voltage becomes 0.87·1.15=1.0 p.u.

## D. System Expansion

Wind farms can be built in successive steps such that the power transmission is initially less than the cable transfer capability. In such scenario, it will be beneficial to be able to operate the cable at both 1.0 p.u. and at a substantially reduced voltage. As an example, consider the situation that the wind farm is being developed in two stages where an initial installation of 150 MW is increased to 300 MW. In this case, it is desirable to be able to achieve high efficiency at both 150 MW and 300 MW installed production.

Fig. 13 shows the cable efficiency as function of installed farm production, with alternative voltage variation ranges with OLTC. It is observed that it is in this case desirable to have a quite large voltage variation range in order to allow a high efficiency also at 150 MW. For instance, allowing 0.6-1.0 p.u. variation in the voltage improves the cable efficiency at 150 MW installed power by 2.5% compared to operation with 1.0 p.u. fixed voltage, and by about 1% at 300 MW installed power. Table II lists the percent voltage variations associated with the ranges in Fig. 13. It is for instance seen that a 0.6-1.0 p.u. voltage variation implies a nominal voltage ratio of 0.8 p.u. with a ±25% regulation.

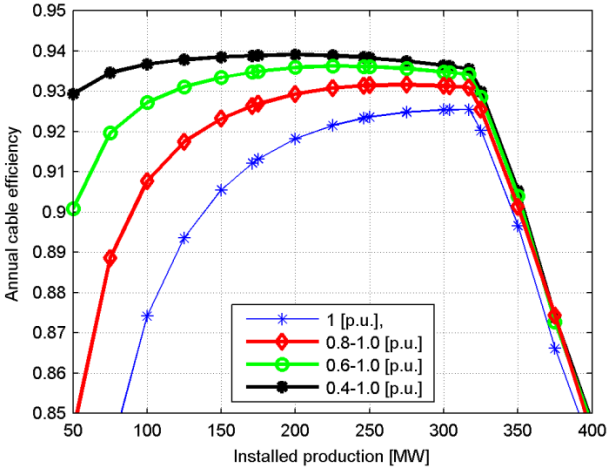

Fig. 13. Annual cable efficiency as function of wind farm maximum instantaneous production for different voltage regulation intervals.

TABLE II.
VOLTAGE REGULATION RANGES IN FIG. 13.

| Range | Nominal voltage | Variation |
|---|---|---|
| 0.8-1.0 p.u. | 0.9 p.u. | ±11.1 % |
| 0.6-1.0 p.u. | 0.8 p.u. | ±25.0 % |
| 0.4-1.0 p.u. | 0.7 p.u. | ±42.9 % |

## VIII. DISCUSSION

### A. Loss Reduction Potential

The results in Sections VI and VII show that there is a significant potential for increased annual cable efficiency and consequently reduced losses. It is important to realize that what appears to be a small increase in efficiency actually represents a large reduction in losses and consequently a large reduction in associated costs. For instance, an increase of the efficiency from 0.92 to 0.93 implies a loss reduction of 12.5%.

Table III summarize the loss reduction potential for some selected cases for the wind farm with high utilization factor. The content in the tables are based on readouts from the presented plots. The reference case for the table is operation of cable at fixed rated voltage (1.0 p.u.).

TABLE III.
SAMPLES OF ANNUAL LOSS REDUCTION POTENTIAL FOR THE 200 KM TRANSMISSION FOR WIND FARM WITH <u>HIGH</u> UTILIZATION FACTOR

| Wind farm rating [MW] | Operation | Annual efficiency improvement | Percent reduction in annual losses |
|---|---|---|---|
| 320 | Variable voltage 0.4-1.0 p.u. | 0.925 → 0.935 | 13% |
| 320 | Variable voltage 0.87 p.u. ±15% | 0.925 → 0.932 | 9% |
| 200 | Fixed voltage 0.8 p.u. | 0.92 → 0.925 | 6% |
| 200 | Variable voltage 0.8 p.u. ±15% | 0.92 → 0.932 | 15% |
| 200 | Variable voltage 0.4-1.0 p.u. | 0.92 → 0.94 | 25% |

### B. Tap-Changer

In this work, it is assumed that the tap-changers have infinitely small steps and that there are no limitations in how often they are allowed to be operated. In reality, there will be a limited number of steps and one will most likely have to restrict how often the tap-changers are operated in order to limit the wear-and-tear. This will give a somewhat smaller reduction in losses but it is not believed to have significant impact on the results since the fluctuations in power production for a windfarm are rather slow.

It is acknowledge that a voltage regulation of 0.4-1.0 p.u. is very high. It was included in the analysis in order to reveal the full potential of voltage regulation. There are however, examples of power transformers in use with a quite large regulation range. One example is the third pole of the Skagerak HVDC connection where the transformer voltage regulation is +30/−10% [9]

### C. Wind Farm Utilization Factor

The calculated results in Section VII demonstrated that allowing regulation of the operating voltage allows substantial improvements to the cable efficiency when taking into account the annual distribution of the wind farm production. That result was for a specific case with high utilization factor. In the case of wind farm with lower utilization factors, the improvements to cable efficiency are even higher. Fig. 14 shows the relative duration of the wind farm annual production of such a case where the utilization factor is 0.35. Compared to the previous result in Fig. 10 that has a utilization factor of 0.46, the average production relative to maximum installed power is lower. Fig. 15 shows the annual cable efficiency associated with this power distribution. As expected, the annual efficiency is with fixed operating voltage lower than in the case of high utilization factor (Fig. 11) whereas the annual efficiency in Figs. 11 and 15 are almost equal when using the 0.4-1.0 p.u. voltage variation. It can therefore be concluded that the value of operating at variable voltage increases with decreasing utilization factor.

Table IV summarizes the loss reduction potential for two cases for the wind farm with low utilization factor. The reference case for the table is operation of cable at fixed rated voltage (1.0 p.u.).

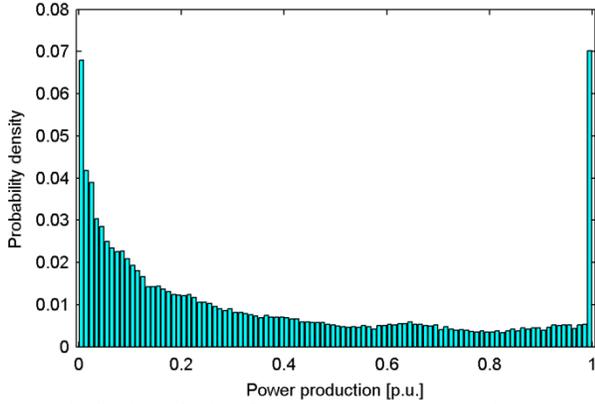

Fig. 14. Distribution of wind farm annual power production in [p.u.] of maximum installed production. Data for wind farm with low utilization factor.

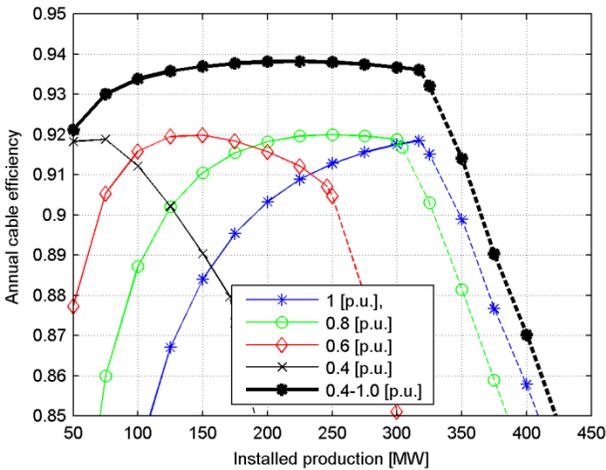

Fig. 15. Annual cable efficiency as function of wind farm maximum instantaneous production. Wind farm with low utilization factor. The dashed traces represent situation with curtailment of production.

TABLE IV.
SAMPLES OF ANNUAL LOSS REDUCTION POTENTIAL FOR THE 200 KM TRANSMISSION FOR WIND FARM WITH LOW UTILIZATION FACTOR

| Wind farm rating [MW] | Operation | Annual efficiency improvement | Percent reduction in annual losses |
|---|---|---|---|
| 320 | Variable voltage 0.4-1.0 p.u. | 0.918 → 0.935 | 21% |
| 200 | Fixed voltage 0.8 p.u. | 0.903 → 0.92 | 18% |

### D. Cable Length and Cable Design Parameters

Most calculated results assumed a cable length of 200 km. This choice was based on the fact that few wind farm installations exist with more than 100 km connection length. It is therefore a need for new operating principles and/or technologies to make AC transmission beyond the 100 km distance viable, and this work is a contribution in that direction.

The analysis also considered one specific cable design. It is clear, however, that the cable electrical per-unit-length parameters ($R$, $L$, $C$) are dependent on the cable design, giving additional instrument to be included in the optimization.

### E. Cable Temperature Variation

The cable is in the analysis represented by a distributed-parameter model to properly take into account the variation of losses along the cable associated with the charging currents. However, it is assumed that the AC resistance is constant along the cable, thereby ignoring the temperature variation along the cable. This assumption will have some influence on the numerical values, but is not believed to have major impact on the relative reduction in losses when operating at optimal voltage. Such temperature variation can be easily included in the analysis by segmentation of the cable.

### F. Other System Losses

This work focuses on the losses in the cable only. It is clear that operation at a reduced voltage will also affect the losses in in other system parts, e.g. the two transformers. However, as transformer losses are low compared to the losses of very long cables, they are not expected to have a significant impact on the conclusions.

### G. Reactive Power Compensation

The analysis has tacitly assumed that the operation of the system is such that the reactive power produced by the cable can be absorbed at both ends. In practice, this implies that about 50% of the reactive power is consumed by the wind farm. This consumption can be achieved using conventional shunt reactors, or by means of controlling the wind turbines. It is emphasized that the shunt reactors do not need to be controlled when the cable operating voltage is adjusted since the cable VAR production and the reactor VAR compensation are both proportional to the square of the operating voltage.

### H. Operating Voltage for Maximum Power Transfer Capability.

In addition to improving cable efficiency, it is also possible to use voltage adaption for extending the maximum useful power transfer capability of a given cable. Using the system model described in Sections II and III together with the cable parameters in Table I, we analyze the maximum power transfer that can be achieved as function of the cable length and the cable operating voltage, without consideration to the cable efficiency. The maximum power transfer is calculated by searching for the wind farm voltage (magnitude factor $\alpha$ and phase angle $\beta$) which maximizes the transmitted power while respecting the current limit in the cable. The permissible voltage variations on the wind farms side are defined in Section III.

Fig. 16 shows with thin lines the maximum power transfer capability as function of the cable length, with the cable operating voltage as parameter (four alternative voltage levels). The solid lines denote the power supplied by the wind farm so that the difference between dashed and solid line represent the cable losses. The thick lines denote the envelope curves that result if one for each length operates at exactly the voltage that maximizes the transmission capacity. It is observed that by reducing the operating voltage it becomes possible to transmit power over longer distances. For instance, with 1.0 p.u. operating voltage the maximum useful cable length is shorter than 270 km. By reducing the operating voltage to 0.6 p.u., the same cable can be used for lengths up to 400 km, although with a reduction in both maximum permissible transmitted power and cable efficiency.

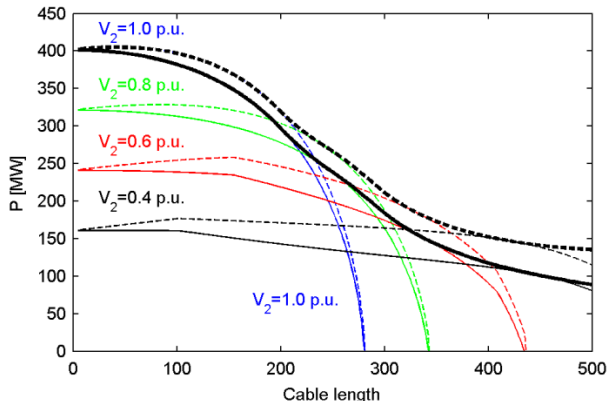

Fig. 16. Thin lines: maximum power transfer capability as function of cable length with (fixed) cable operating voltage (onshore side) as parameter. Dashed lines: produced power at wind farm; solid lines: power delivered to transformer on shore side. Thick lines: ditto result with use of optimal operating voltage. All curves are for the cable with parameters as in Table I.

## IX. Conclusions

This study considers a wind farm HVAC transmission system where a cable connects a wind farm to the onshore grid via two transformers. The cable efficiency is analyzed using a detailed cable model based on distributed electrical parameters. From the analysis, the following conclusions are reached:

1. For a given cable length, the maximum attainable cable efficiency is independent of the cable operating voltage.
2. The operating voltage affects the power transmission level at which the maximum cable efficiency is attained. The maximum efficiency appears at lower power levels when operating voltage is reduced.
3. The cable efficiency can be increased if tap-changers are used to adjust the operating voltage according to the variations in the instantaneous wind power production levels. Calculations for a 200 km cable connecting a 320 MW windfarm showed that loss reduction of 9% is achievable by simply using a ±15% voltage regulation of the two transformers.
4. Usage of an even higher regulation leads to further improvements in the cable efficiency. If voltage can be varied between 0.4 and 1.0 p.u. one can achieve a loss reduction of 13% for the same wind farm.
5. The benefit of variable transmission voltage is highest for wind farms having a low utilization factor. A loss reduction of 21% was demonstrated for a 200 km/320 MW windfarm with low utilization factor when operated with variable voltage between 0.4 and 1.0 p.u.
6. Usage of a reduced operating voltage can also be used as a means of increasing the maximum transmission length for a given cable, although the permissible level of the transmitted power is reduced compared to short lengths.

The results presented here are relevant for those who are planning and engineering wind farms as well as for those optimizing cable designs for a given plant. The proposed methodology has the advantage that it can be realized without introducing new technology that needs to be developed or qualified.

## XI. Biographies

**Bjørn Gustavsen** (M'94–SM'2003–F'2014) was born in Norway in 1965. He received the M.Sc. degree and the Dr.Ing. degree in Electrical Engineering from the Norwegian Institute of Technology (NTH) in Trondheim, Norway, in 1989 and 1993, respectively. Since 1994, he has been working at SINTEF Energy Research where he is currently Chief Scientist. His interests include simulation of electromagnetic transients and modeling of frequency dependent effects. He spent 1996 as a Visiting Researcher at the University of Toronto, Canada, and the summer of 1998 at the Manitoba HVDC Research Centre, Winnipeg, Canada. He was a Marie Curie Fellow at the University of Stuttgart, Germany, August 2001–August 2002. He is convenor of CIGRE JWG A2/C4.52.

**Olve Mo** was born in Norway in 1965. He received the M.Sc. degree and the Dr.Ing. degree in Electrical Engineering from the Norwegian Institute of Technology (NTH) in Trondheim, Norway, in 1988 and 1993, respectively. Since 2014, he has been working at SINTEF Energy Research where he is currently Research Scientist. His interests are power converters and energy storage in power systems, including wind farms and vessel propulsion systems. His former experience include eight years at Marine Cybernetics, developing simulator based testing technology for control systems software on advanced offshore vessels, five years former employment at SINTEF Energy Research, five years at Siemens in Trondheim working with switchgear deliveries to the oil and gas industry, and two years as associate professor at the High Voltage Section at NTNU.